\documentclass[12pt,twocolumn, aps,nofootinbib,preprintnumbers,superscriptaddress,numberedappendix]{openjournal}
\usepackage{natbib}

\usepackage{newtxtext,newtxmath}

\usepackage[T1]{fontenc}

\usepackage[colorlinks=true, allcolors=blue]{hyperref}

\DeclareRobustCommand{\VAN}[3]{#2}
\let\VANthebibliography\thebibliography
\def\thebibliography{\DeclareRobustCommand{\VAN}[3]{##3}\VANthebibliography}

\usepackage{graphicx}	
\usepackage{amsmath}	
\usepackage{soul}
\usepackage{microtype}
\usepackage{pythonhighlight}
\usepackage{xspace}
\usepackage{makecell}
\usepackage{pmboxdraw} 

\newcommand{\galsim}[0]{\textsc{GalSim}\xspace}

\newcommand{\astropy}[0]{\texttt{astropy}\xspace}
\newcommand{\surveycodex}[0]{\textsc{SurveyCodex}\xspace}
\newcommand{\scarlet}[0]{\textsc{Scarlet}\xspace}
\newcommand{\catsim}[0]{CatSim\xspace}
\newcommand{\btk}[0]{\texttt{BTK}\xspace}

\newcommand{\rfig}[1]{Figure~\ref{fig:#1}}
\newcommand{\rtab}[1]{Table~\ref{tab:#1}}

\newcommand{\rsec}[1]{Section~\ref{sec:#1}}
\newcommand{\rapp}[1]{Appendix~\ref{app:#1}}

\newcommand{\orcidauthor}[3]{\author{\href{http://orcid.org/#1}{#2$^{#3}$}}}

\shorttitle{Blending Toolkit}
\shortauthors{Mendoza and Torchylo, et al.}
\begin{document}


\title{The Blending ToolKit: \\ A simulation framework for evaluation of galaxy detection and deblending}

\orcidauthor{0000-0002-6313-4597}{Ismael Mendoza}{1, *, \star}
\orcidauthor{0009-0008-4259-8254}{Andrii Torchylo}{2, \dag, \star}
\orcidauthor{0009-0003-0169-266X}{Thomas Sainrat}{3, 4, \ddag}
\orcidauthor{0000-0002-5068-7918}{Axel Guinot}{4, 5}
\orcidauthor{0000-0001-7387-2633}{Alexandre Boucaud}{4}
\orcidauthor{0000-0002-0849-7707}{Maxime Paillassa}{6}
\orcidauthor{0000-0001-8868-0810}{Camille Avestruz}{1,7}
\orcidauthor{0000-0001-9431-3806}{Prakruth Adari}{8}
\orcidauthor{0000-0002-5592-023X}{Eric Aubourg}{9}
\orcidauthor{0000-0001-8287-7444}{Biswajit Biswas}{4}
\orcidauthor{0000-0001-8207-5556}{James Buchanan}{10}
\orcidauthor{0000-0001-5004-3736}{Patricia Burchat}{11}
\orcidauthor{0000-0003-4480-0096}{Cyrille Doux}{12}
\orcidauthor{0000-0002-2704-5028}{Remy Joseph}{13}
\orcidauthor{0000-0003-0443-8221}{Sowmya Kamath}{2}
\orcidauthor{0000-0002-8676-1622}{Alex I. Malz}{5}
\orcidauthor{0009-0005-7923-054X}{Grant Merz}{14}
\orcidauthor{0000-0001-7964-9766}{Hironao Miyatake}{15, 16, 17}
\orcidauthor{0000-0002-9641-4552}{C\'ecile Roucelle}{4}
\orcidauthor{0000-0002-5596-198X}{Tianqing Zhang}{5, 18, 19}
\author{the LSST Dark Energy Science Collaboration}

\thanks{$^*$ Corresponding Author: \href{mailto:imendoza@umich.edu}{imendoza@umich.edu}.}
\thanks{\dag~Corresponding Author: \href{mailto:torchylo@stanford.edu}{torchylo@stanford.edu}.}
\thanks{\ddag~Corresponding Author: \href{mailto:thomas.sainrat@iphc.cnrs.fr}{thomas.sainrat@iphc.cnrs.fr}.}
\thanks{$\star$~These authors contributed equally.}
\thanks{For more details on author contributions, see Section~\ref{sec:contributions}.}

\affiliation{$^{1}$Department of Physics, University of Michigan, Ann Arbor, MI 48109, USA}
\affiliation{$^{2}$Department of Physics, Stanford University,
Stanford, CA 94305, USA}
\affiliation{$^{3}$Université de Strasbourg, CNRS, IPHC UMR 7178, F-67000 Strasbourg, France}
\affiliation{$^{4}$Université Paris Cité, CNRS, AstroParticule et Cosmologie, F-75013, Paris, France}
\affiliation{$^{5}$McWilliams Center for Cosmology, Department of Physics, Carnegie Mellon University, Pittsburgh, PA 15213, USA}
\affiliation{$^{6}$Department of Physics, Graduate School of Science, Nagoya University, Furo-cho Chikusa-ku, Nagoya 464-8602, Japan}
\affiliation{$^{7}$Leinweber Center for Theoretical Physics, University of Michigan, Ann Arbor, MI 48109, USA}
\affiliation{$^{8}$Department of Physics and Astronomy, Stony Brook University, Stony Brook, NY 11794-3800, USA}
\affiliation{$^{9}$Université Paris Cité, CNRS, CEA, AstroParticule et Cosmologie, F-75013, Paris, France}
\affiliation{$^{10}$Physics Division, Lawrence Livermore National Laboratory, Livermore, CA 94306, USA}
\affiliation{$^{11}$Kavli Institute for Particle Astrophysics and Cosmology, Department of Physics, Stanford University, Stanford, CA 94305, USA}
\affiliation{$^{12}$Univ. Grenoble Alpes, CNRS, Grenoble INP, LPSC-IN2P3, 38000 Grenoble, France}
\affiliation{$^{13}$The Oskar Klein Centre, Department of Physics, Stockholm University, Albanova University Center, Stockholm, SE-10691, Sweden.}
\affiliation{$^{14}$Department of Astronomy, University of Illinois at Urbana-Champaign, 1002 West Green Street, Urbana, IL 61801, USA}
\affiliation{$^{15}$Kobayashi-Maskawa Institute for the Origin of Particles and the Universe (KMI), Nagoya University, Nagoya, 464-8602, Japan}
\affiliation{$^{16}$Institute for Advanced Research, Nagoya University, Nagoya, 464-8601, Japan}
\affiliation{$^{17}$Kavli Institute for the Physics and Mathematics of the Universe (WPI), The University of Tokyo Institutes for Advanced Study (UTIAS), The University of Tokyo, Chiba 277-8583, Japan}
\affiliation{$^{18}$Department of Physics and Astronomy and PITT PACC, University of Pittsburgh, Pittsburgh, PA 15260, USA}
\affiliation{$^{19}$SLAC National Accelerator Laboratory, 2575 Sand Hill Road, Menlo Park, CA 94025, USA}


\begin{abstract}

We present an open source Python library for simulating overlapping (i.e., \textit{blended}) images of galaxies and performing self-consistent comparisons of detection and deblending algorithms based on a suite of metrics.
The package, named \textit{Blending Toolkit} (\btk), serves as a modular, flexible, easy-to-install, and simple-to-use interface for exploring and analyzing systematic effects related to blended galaxies in cosmological surveys such as the Vera Rubin Observatory Legacy Survey of Space and Time (LSST). 
\btk has three main components: (1) a set of modules that perform fast image simulations of blended galaxies, using the open source image simulation package \galsim; (2) a module that standardizes the inputs and outputs of existing deblending algorithms; (3) a library of deblending metrics commonly defined in the galaxy deblending literature. 
In combination, these modules allow researchers to explore the impacts of galaxy blending in cosmological surveys. 
Additionally, \btk provides researchers who are developing a new deblending algorithm a framework to evaluate algorithm performance and make principled comparisons with existing deblenders.
\btk includes a suite of tutorials and comprehensive documentation. The source code is publicly available on GitHub at \url{https://github.com/LSSTDESC/BlendingToolKit}. 
\end{abstract}

\maketitle

\section{Introduction}

Modern ground-based cosmological surveys, including the Dark Energy Survey \citep{des2016des}, the Hyper Suprime-Cam Subaru Strategic Program \citep{hiroaki2018hsc}, and the upcoming Rubin Observatory Legacy Survey of Space and Time (LSST, \cite{ivezic2019lsst}) will detect an unprecedented number of galaxies and have the potential to significantly improve our understanding of the Universe. 
As the sensitivity of these surveys increases, so does the density of detectable light sources (stars and galaxies), which in turn increases the rate at which sources overlap with each other in images.
This phenomenon, known as \textit{blending}, is a major potential source of systematic bias and uncertainty for modern surveys \citep{melchior2021challenge}; blending undermines our ability to identify individual sources in a given image \citep{dawson2016ambiguous,troxel2023joint} and disentangle their properties, such as flux and shape, from those of neighboring sources \citep{dawson2016ambiguous,hoekstra2017shape,euclid2019undetected}.

In LSST, it is predicted that about $60\%$ of galaxies will be blended to some degree (at full-depth, i.e. $r\simeq27.2$) \citep{sanchez2021effects}, $51\%$ of galaxies will be overlapping with sources that can be detected individually (at depth $r\simeq24$) \citep{eckert2020noise}, and that between $15\%$ and $30\%$ of galaxies in $i$-band images will be part of \textit{unrecognized blends} (at depth $i\simeq 27$ and $i \simeq26$ respectively) \citep{dawson2016ambiguous,troxel2023joint} -- i.e., blends that are detected as a single object, but actually correspond to two or more sources.
For HSC, it was estimated that about $58\%$ of the detected objects in the \textit{i}-band wide survey are blended sources (at depth $r\simeq26$) \citep{bosch2017hsc_pipeline}.
Since a large fraction of the galaxies detected in modern surveys will be impacted by blending, blended objects cannot be discarded without losing a significant amount of statistical power \citep{chang2013number,chang2015erratum} and inducing selection bias in the cosmological probes \citep{hartlap2011bias}.
Thus, we need an accurate understanding of the systematic errors that blending contributes to our measurements, and robust algorithms to mitigate these errors.

Blending impacts the two most important static probes of dark energy for ground-based cosmological surveys: weak gravitational lensing and galaxy clusters. 
Weak gravitational lensing uses the fact that observed galaxy shapes trace the underlying dark matter distribution \citep{huterer2002weak,kilbinger2015cosmology}. 
In principle, blending impacts the three types of two-point statistics (cosmic shear, galaxy-galaxy lensing, and galaxy clustering) commonly used in weak lensing analyses, although its effect on cosmic shear has been most widely studied.
\textit{Cosmic shear} is the relatively small, approximately linear distortion of shapes by weak lensing from large-scale mass distributions.
Blending can introduce several types of potential bias in measured shear: selection bias due to detection inefficiency for blended galaxies \citep{kaiser2000shear,bernstein2002shapes,hirata2003shear,dawson2016ambiguous}; shape bias for blended galaxies \citep{dawson2016ambiguous,hoekstra2017shape,samuroff2018des_y1,euclid2019undetected}; and  
increased pixel-noise bias and reduced statistical sensitivity \citep{sanchez2021effects}. Additionally, blending impacts galaxy photometry \citep{huang2018photometric,boucaud2020coindeblend}, and even causes (rare) catastrophic outliers in photometry measurements \citep{everett2022transfer}.
Because blending will occur between galaxies at different redshifts \citep[e.g.,][]{dawson2016ambiguous}, photometric redshift (photo-$z$) estimates, which are an essential ingredient of weak lensing cosmological analyses, are impacted by blending \citep{mandelbaum2018precision}. 
Algorithms are already in development for measuring photo-$z$ in the presence of blends \citep{jones2019bpz,jones2019gmm}. 
However, more studies are needed to fully understand the impact of blending on photo-$z$ estimation for cosmic shear analyses. 

Quantities of interest in cluster cosmology, including total cluster mass and redshifts of member galaxies, provide another important cosmological probe \citep[e.g.,][]{planck2016clusters,lesci2022amico,DESY1cluster,sunayama2023optical,bocquet2024spt,weighinggiantsiv}, as they describe the high density peaks in the mass power spectrum, providing complementary information to the weak lensing results.
Galaxy clusters host a particularly high density of galaxies, which increases the rate of blending \citep{liang2022photometry}; therefore, blending has the potential to bias these important measurements. 
The best known way to calibrate the total cluster mass is by measuring the weak lensing distortion of background source galaxies \citep[e.g.,][]{metzler1999mass}. 
Recent studies show that \textit{recognized} galaxy blends -- i.e., blends in which each galaxy member is accurately identified as an individual galaxy -- in clusters could bias the weak lensing profile ($\Delta \Sigma (r)$) by about $20\%$ \citep{ramel2023blending}, likely due to shape misestimation.
One of the main sources of bias in estimating the cluster mass is the misidentification of cluster members as background galaxies \citep[e.g.,][]{varga2019contamination}. The classification accuracy is strongly dependent upon precise photo-$z$ estimates for cluster member and background galaxies \citep[e.g.,][]{esteves2024copacabana}. 
Blending has the potential to increase the rate of misidentification in two ways: (1) cluster and background galaxies could be blended to the extent that they become unrecognized blends, at which point contamination becomes unavoidable; (2) blending from faint sources could impact the observed colors of galaxies \citep{gruen2019icl}, which would bias estimates of photo-$z$s.

Given the plethora of complicated biases blending can induce in cosmology probes, it is essential to leverage simulations to understand the impact of blending on these probes and to calibrate blending related biases in analyses \citep{melchior2021challenge}.
A widely used package for generating simulated galaxy images is \galsim \citep{galsim2015}, which can generate galaxy images for a range of galaxy profiles, point spread functions (PSF), and noise levels. \galsim has been used to calibrate blending biases for weak lensing \citep[e.g.,][]{maccrann2022desy3,dalal2023hsc_y3_cosmo}, and forms the basis for several large-scale simulations of cosmological surveys. 
One such simulation emulating LSST observations is the LSST DESC second data challenge (DC2) Simulated Sky Survey \citep{lsst2021dc2}, which uses a full end-to-end approach -- starting from N-body cosmological simulations and outputting catalogs based on forward-modeled astronomical images processed by the LSST Science Pipelines. 
DC2 has already been used to support various blending studies in the context of LSST \citep{buchanan2022gp,ramel2023blending}.
Finally, another powerful tool is to use multiple surveys with different angular resolutions, such as a ground-based and a space-based survey, to characterize the rate of blending \citep{mandelbaum2018hsc,troxel2023joint} or improve deblending \citep{joseph2021joint}.

Another approach to mitigating the blending problem is the development and application of \textit{deblenders}: algorithms designed to measure the individual properties of each galaxy in a blend. 
In principle, an ideal deblender would allow the recovery of these individual properties and the corresponding catalogs could be analyzed ignoring blending altogether. 
However, in practice, detection and deblending can fail in various ways: blends may not be identified (unrecognized blends), and the deblending process might infer galaxy properties inaccurately depending on the severity of the blend. The former effect can introduce an additional source of systematic error in cosmic shear measurements \citep{nourbakhsh2022blending}.
SourceExtractor (SExtractor) is an early deblender that uses a thresholding technique for object detection and a multiple isophotal analysis for deblending \citep{bertin1996sextractor}. 
More modern deblenders use non-parametric constrained optimization techniques \citep{joseph2016multi,melchior2018scarlet}, such as the \scarlet deblender, or machine learning techniques \citep{arcelin2021vae,hansen2022deblending,liu2023variational,merz2023deepdisc,sampson2024scarlet2,biswas2024madness}.

Despite the range of available simulation packages and deblenders, there is currently no standard framework dedicated to creating customizable simulations of blended galaxies, or comparing the performance of different deblending algorithms using consistent simulated galaxy densities, fluxes, noise levels, image resolutions, etc., and consistent deblending metrics. 
To address this limitation, we introduce the \textit{Blending Toolkit} (\btk), a modular Python-based software package that leverages \galsim to generate reproducible and customizable astronomical images of galaxy blends under the observing conditions of several dark energy surveys. 
\btk also includes a module standardizing existing deblenders such that their outputs can be directly compared against one another when applied to the same images. 
Finally, our package includes a growing library of deblending metrics to enable evaluation of the performance of deblenders. 

We will first describe the design philosophy for \btk in \rsec{development}. Then, we describe the software modules comprising \btk in detail \rsec{structure}, followed by an outline of the existing Jupyter Notebook tutorials and documentation in \rsec{tutorials}. We provide a brief summary and outlook for the future in \rsec{summary}.

\section{Development Approach}
\label{sec:development}

The original motivation for \btk was to build a framework that could bring together the various simulation and deblending efforts within the LSST Dark Energy Science Collaboration (DESC) Blending Working Group and the broader cosmology community. 
We learned of a number of ideas for deblenders being developed; however, it was difficult to assess their relative performance compared to existing state-of-the-art deblenders, or their strengths and weaknesses. This was mainly because the majority of deblender studies used a different set of astronomical images (or simulations) and metrics to evaluate their deblenders. 
In order to make progress in deblender development, a way to assess and benchmark existing deblenders consistently was needed, and this is one of the main purposes of \btk. 

\btk is designed to be:  
\begin{itemize}
    \item \textbf{Modular}: The simulation, deblending, and metric components of \btk can be connected together to form a full analysis pipeline, as demonstrated in the tutorials available for \btk (see \rsec{tutorials}); however, these components are also designed to be usable independently.

    \item \textbf{Customizable}: Users can customize the various \btk modules to fit their desired blending analysis. For example, the simulations of blended galaxies within \btk contain many options for changing the configuration of galaxy blends (e.g., the rate of blending) or the observing conditions (e.g., noise level or PSF). Comprehensive documentation and examples make it clear how to change the various options within the package. 

    \item \textbf{Parallelizable and Reproducible}: Every computationally intensive process in \btk (e.g., producing simulations or running deblenders) can be split across multiple CPUs to improve efficiency. We carefully ensured that all simulations produced by \btk are reproducible given a single random seed. Randomness is propagated using a \texttt{numpy.random.SeedSequence} object during multiprocessing to guarantee said reproducibility.

    \item \textbf{Easy-to-Install}: \btk only has a few Python dependencies, the main one being \galsim. This ensures that the package is easily installable with a single \texttt{pip} command.
\end{itemize}

\section{Structure}
\label{sec:structure}

\begin{figure}
    \centering
    \includegraphics[width=0.47\textwidth]{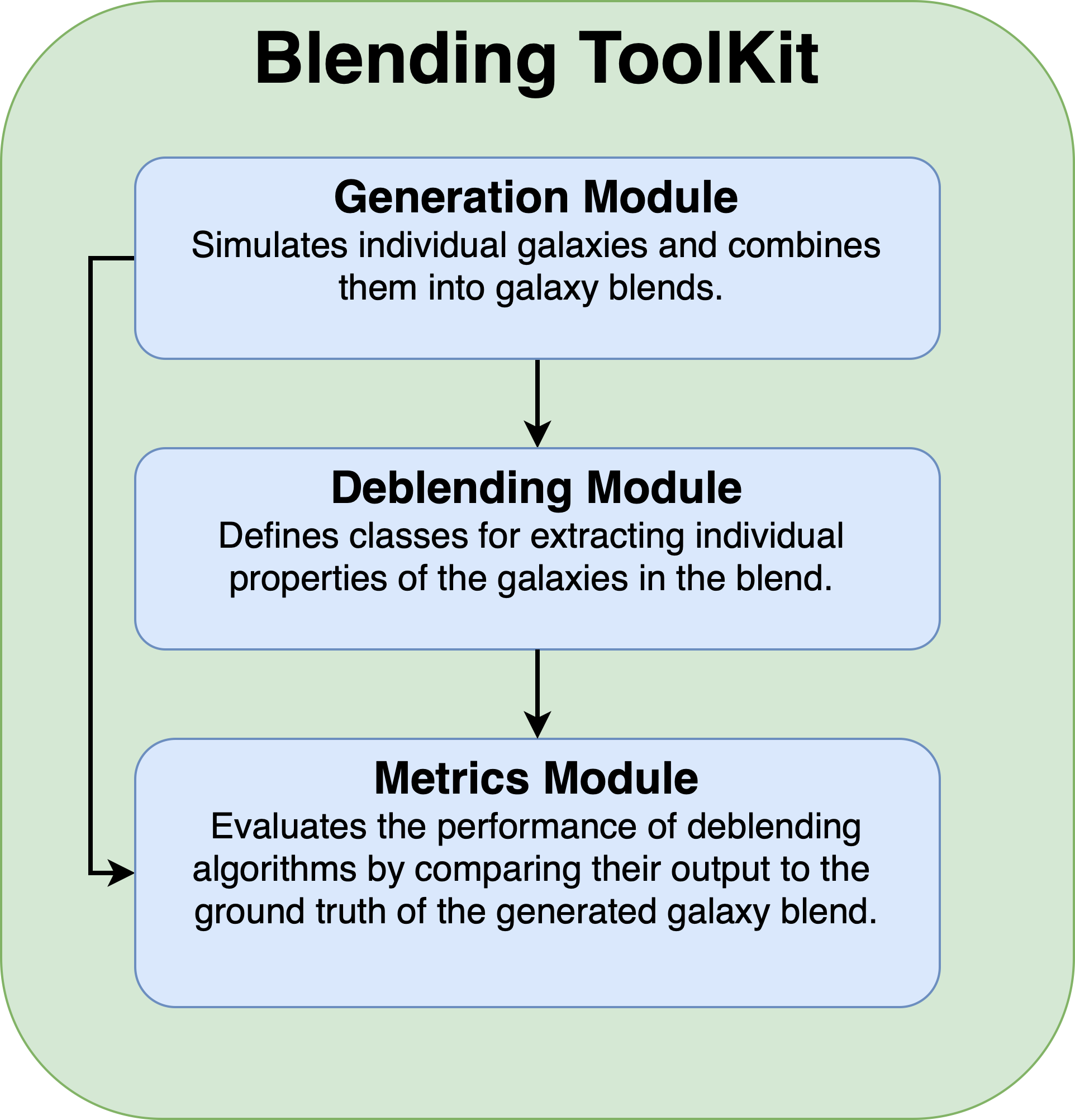}
    \caption{
    \textbf{Overall \btk structure.} The software package consists of three main modules: \textit{Generation}, \textit{Deblending}, and \textit{Metrics}. Each arrow indicates the data flow between modules. 
    }
    \label{fig:pipeline}
\end{figure}

The overall structure of \btk is summarized in \rfig{pipeline}.
The \textit{Generation} module provides the simulated data containing blended galaxy images, isolated galaxy images, and ground truth properties of galaxies in the blend. 
Then, with the \textit{Deblending} module, detection or deblending algorithms can be applied to the generated images of galaxy blends. Users can choose to implement their own algorithm or use one of the existing ones in \btk.
Finally, in the \textit{Metrics} module, users can evaluate the performance of their detection or deblending algorithms by comparing their output to the ground truth data.
This module also contains functionality to perform measurements on the simulated images, including photometry and galaxy shapes, which makes it possible to measure the impact of blending on relevant physical quantities.

\subsection{Generation Module}
\label{sec:generation}

The Generation module renders simulations of galaxy blends that can be used as an input to deblenders for their evaluation. Specifically, this module provides a \texttt{DrawBlendsGenerator} class that outputs galaxy blend images in batches i.e., a small collection of objects that can fit in computer memory. This class is implemented as a python generator object, which at each iteration outputs the galaxy blend images, as well as auxiliary galaxy images and truth galaxy catalogs. This generator class also supports multiprocessing to increase the speed of the generation procedure. 

\begin{figure*}
    \centering
    \includegraphics[width=0.75\textwidth]{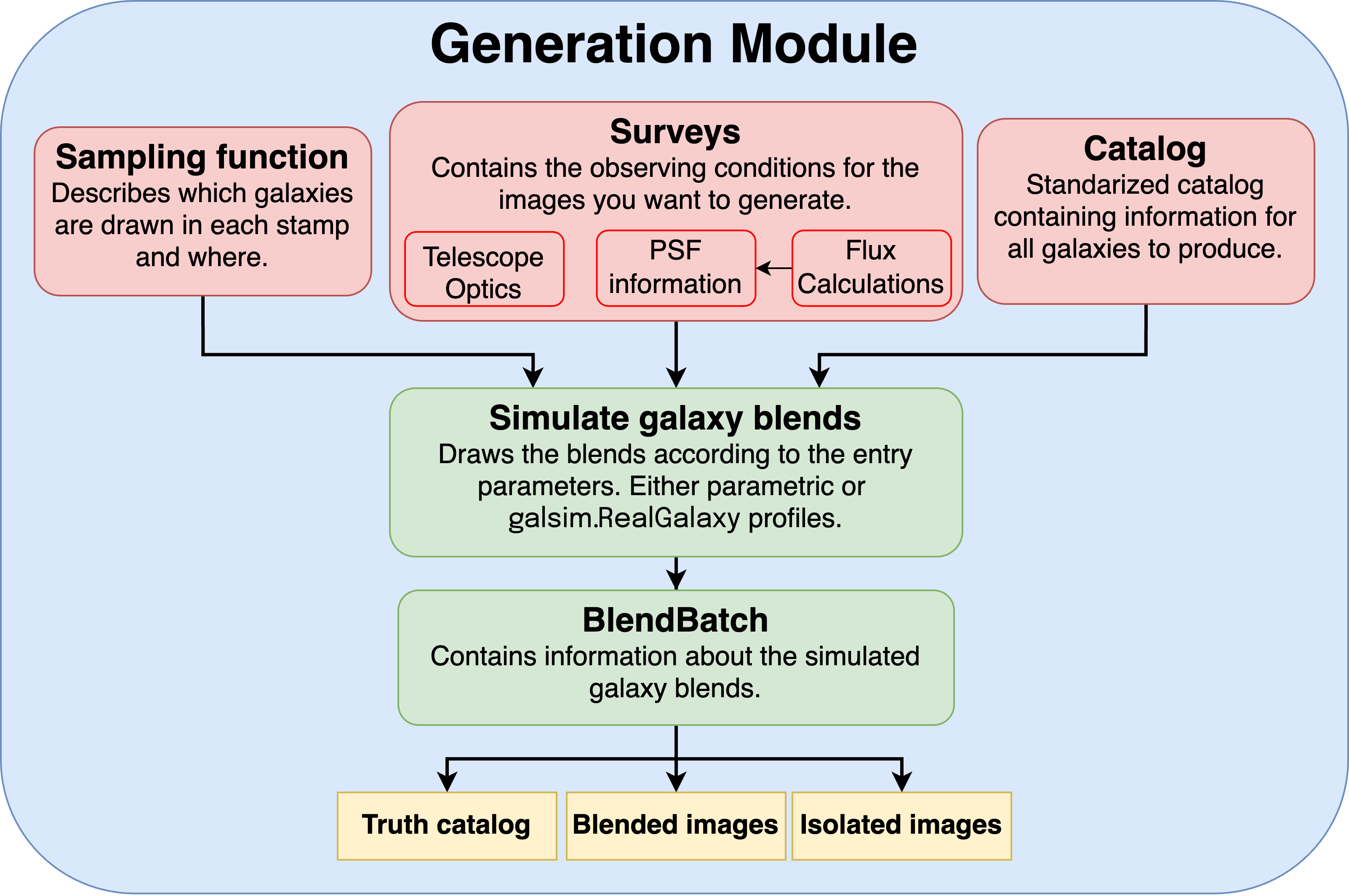}
    \caption{
        \textbf{Generation Module of \btk}. This module contains the functionality and configurations for creating galaxy simulations in \btk. The \texttt{sampling\_function} submodule encodes how galaxies are organized into blends. The \texttt{surveys} submodule specifies the observing conditions. Finally, the \texttt{catalog} submodule encodes the properties of each individual galaxy.
        The red blocks are intended to be easily customized by the user, while green blocks are intended to be used `as-is' and can only be modified by advanced users. Yellow blocks indicate the outputs of a given module. See 
        Section~\ref{sec:generation} for more details on this module.
    }
    \label{fig:generation}
\end{figure*}

Simulated galaxy blends are produced by simulating individual galaxy images with \galsim according to the provided galaxy properties and specified survey observing conditions. 
Then, these individual galaxy images are combined into a single galaxy blend image, and noise is added. 
Additionally, the \texttt{DrawBlendsGenerator} keeps track of the true properties of every galaxy in each blend, and these properties are returned alongside the galaxy blend images and the (noiseless) images of each individual galaxy. These additional outputs are helpful when evaluating or training deblenders, as is the case for ML-based algorithms. For convenience and validation, the output is handled by the \texttt{BlendBatch} class. 
An outline of the overall structure of this module is presented in Figure~\ref{fig:generation}. 

The user can customize the blend generation procedure by modifying the following:
\\[5pt]
\textbf{Catalog:} The \texttt{Catalog} class standardizes the input information about the galaxies that will be used for rendering. 
The underlying structure of the catalog is an \astropy table  \citep{astropy:2013,astropy:2018,astropy:2022}, where each row contains information about a given galaxy's properties such as ellipticity, magnitude in different bands, and size.
\btk currently supports two types of catalog formats: \catsim's parametric catalogs \citep{catsim2014} and \galsim's COSMOS catalogs \citep{mandelbaum2012cosmos_dataset}. Each of these types of catalogs has its dedicated class in \btk. 

The \texttt{CatsimCatalog} class used in \btk uses an \astropy table containing parameters for simulating galaxies based on a three component model: a bulge, a disk and an AGN\footnote{Some galaxies have an Active Galactic Nucleus, which means that their nucleus is much brighter than usual, sometimes as much as the rest of the galaxy.}.
Each galaxy has an overall total flux, each component has an associated flux fraction, and the bulge and disk components have independent sizes and ellipticities. All galaxy properties except for the flux are fixed across all LSST bands.
The input catalog for this class follows the format used by \catsim \citep{catsim2014}, which is the earliest version of an end-to-end LSST simulation framework. \btk provides a sample catalog of approximately $85000$ galaxies taken from the original \catsim simulation\footnote{\url{https://www.lsst.org/scientists/simulations/catsim}}.
The galaxy properties from the \catsim catalog were designed to match the key observables in the LSST survey, and are taken from previous studies on galaxy formation and SDSS observations \citep{bruzual2003stellar, de2006formation, gonzalez2009testing}. 

Alternatively, the image-based COSMOS catalog \citep{mandelbaum2012cosmos_dataset} can be used to generate realistic galaxy profiles. The COSMOS catalog consists of about $80,000$ galaxies extracted from the Hubble Space Telescope (HST) COSMOS survey \citep{koekemoer2007cosmos}, which are used for the base catalog in \galsim\citep{galsim2015}.
Currently, the \texttt{CosmosCatalog} class within \btk enables using the COSMOS catalog to render \btk blends by combining multiple \galsim \texttt{RealGalaxy} profiles that were constructed from HST images.\footnote{For more details see \url{https://galsim-developers.github.io/GalSim/_build/html/real_gal.html}}
When using COSMOS galaxies in a multi-band survey within \btk, the same \texttt{F814W} magnitude is used for all bands, though accounting for the different filters' zeropoints and exposure times. However, a user may also provide multi-band information, which will then be used by \btk to provide an independent flux in each band. Importantly, this functionality does not account for color gradients within a galaxy profile.
\\[5pt]
\textbf{Sampling Function}: The sampling functions are customizable functions that select galaxies from a large catalog to form a given output blend, as well as their relative location within the blend. These functions are designed to be flexible, for example, one of the default sampling functions in \btk randomly selects pairs of galaxies in a catalog and outputs 2-galaxy blends with the brightest galaxy fixed to the center of the stamp, while the dimmer galaxy is a uniformly random distance from the central galaxy. 
Most sampling functions currently implemented support various input parameters that control the degree of blending and total number of galaxies in the galaxy blends produced. 
Users are additionally encouraged to create their own sampling functions depending on their scientific needs. One of the tutorials, \texttt{01-advanced-generation}, detailed in \rsec{tutorials}, provides guidelines and examples on how to achieve this.
\\[5pt]
\textbf{Survey:} The \texttt{Survey} class contains all information of a given astronomical survey, including the telescope optics, which guarantee that generated images approximately reproduce the observing conditions of the ones observed with that particular survey.

To organize this information, \btk uses \surveycodex\footnote{\url{https://github.com/LSSTDESC/surveycodex}}, a tiny Python library with referenced survey parameters for the current main galaxy surveys as well as their filters. 
For more information about \surveycodex and currently supported surveys, see \rapp{surveycodex}. 
Importantly, the \texttt{Survey} class also contains the point-spread function (PSF) that is applied to galaxies, which can also be easily specified and changed by the user. 
\\[5pt]
\indent Once the catalog, sampling function, and survey are specified, simulated data in the form of \texttt{BlendBatch} objects can be easily generated using the \texttt{DrawBlendsGenerator} class.

\begin{figure}
    \centering
    \includegraphics[width=0.46\textwidth]{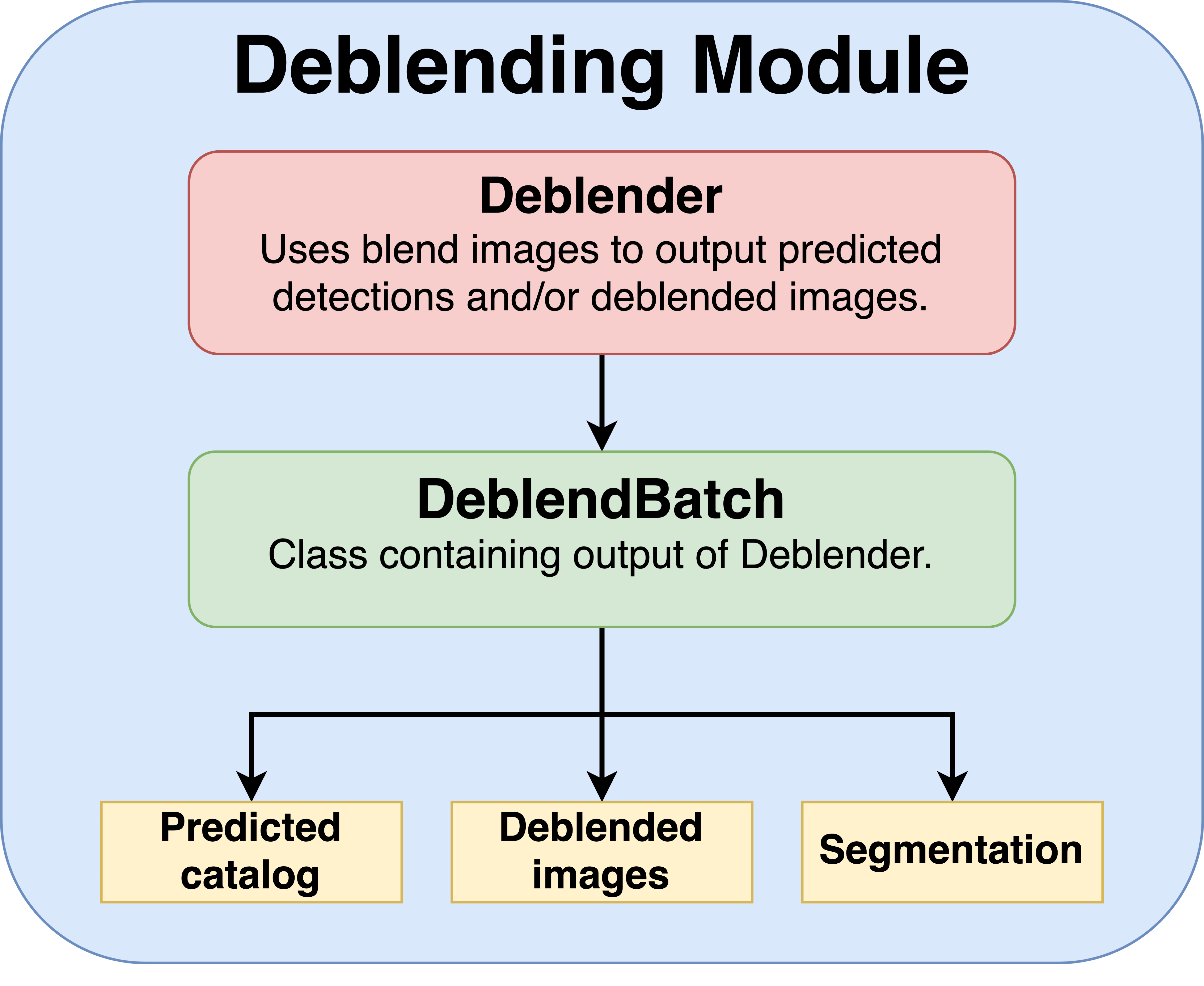}
    \caption{
    \textbf{Deblending Module of \btk}. This module contains a standardized library of deblenders that can be used within \btk. The module is designed to allow users to run their preferred detection or deblending algorithm on simulated data from the Generation module of \btk. Additionally, users can implement a new algorithm by creating a new subclass of the \texttt{Deblender} class. The output of \texttt{Deblender} in \btk is a \texttt{DeblendBatch} object, which can contain a predicted catalog, deblended images, and segmentation.
    See Section~\ref{sec:deblend} for more details on this module.
    }
    \label{fig:deblenders}
\end{figure}

\begin{figure*}
    \centering
    \includegraphics[width=0.75\textwidth]{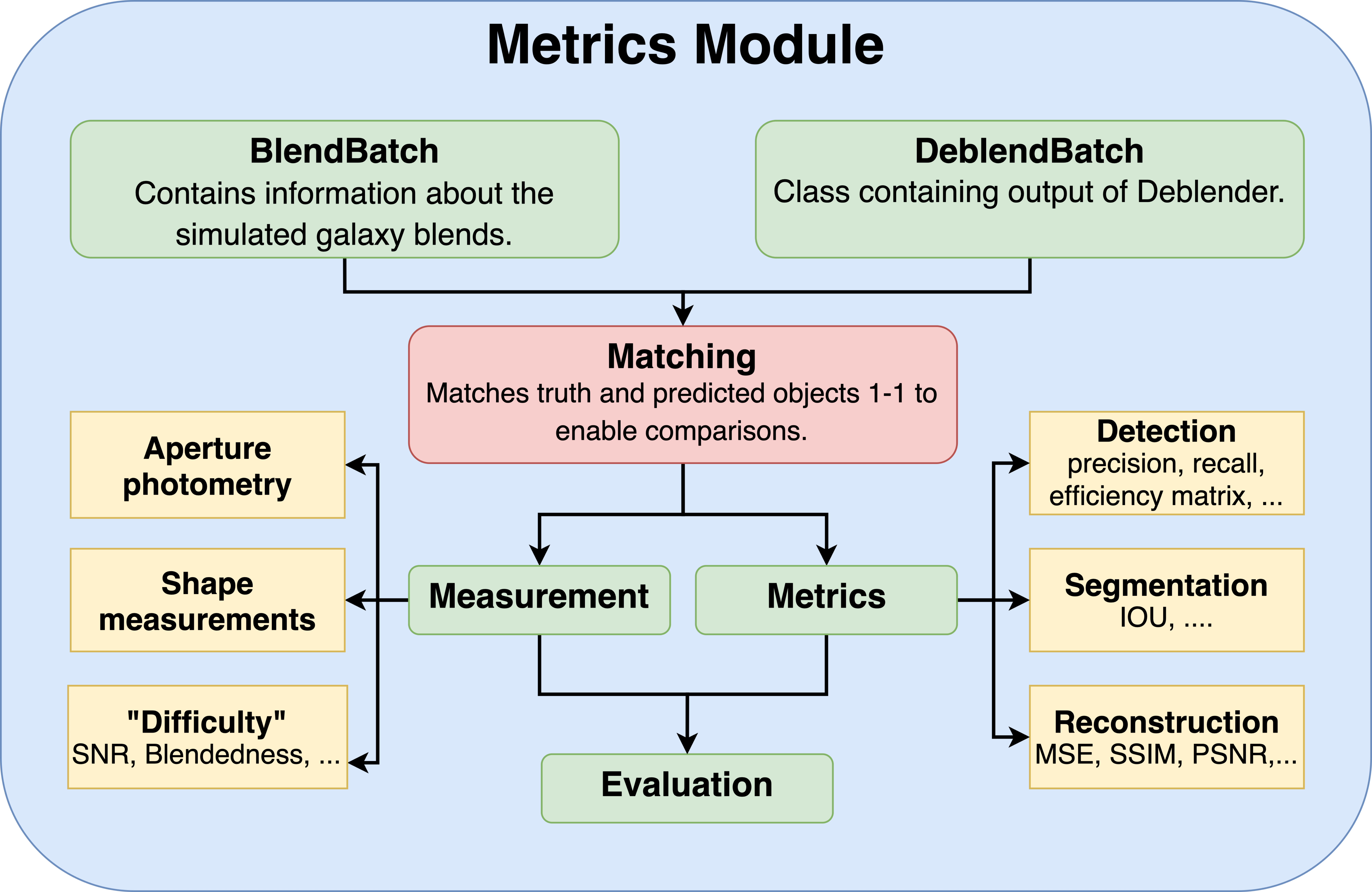}
    \caption{
    \textbf{Metrics module of \btk}. This module uses the outputs of the generation module (the ground truth) and the deblender to evaluate the quality of the deblending. Before comparison, the \texttt{matching} module is used to assign true galaxies to detected galaxies and enable the evaluation of residuals. The matched outputs are subjected to various metrics divided in three categories: \textit{detection}, \textit{segmentation}, and \textit{reconstruction}. 
    Finally, the \texttt{measurement} module uses the images or catalogs from \texttt{BlendBatch} or \texttt{DeblendBatch} to compute physical galaxy quantities for analysis including ellipticities, signal-to-noise-ratio (SNR), and blendedness.
    See Section~\ref{sec:metrics} for more details on this module.
    }
    \label{fig:metrics}
\end{figure*}

\subsection{Deblending}
\label{sec:deblend}

In the context of \btk, we refer to a \textit{deblender} as any algorithm that takes as input images of galaxy blends and attempts to output properties of the individual sources (e.g., centroid location) or a reconstruction of the individual galaxies composing the blend.

One of the main goals of \btk is to allow users to test their preferred detection or deblending algorithm in a controlled setting. 
We provide a framework that enables running such algorithms directly on the \texttt{BlendBatch} output of the Generation module (\rsec{generation}). 
The results can then be analyzed by the user separately or compared with other algorithms using a set of \btk built-in blending metrics, as described in \rsec{metrics}.
An outline of overall structure of the deblending module is presented in \rfig{deblenders}.

Our deblender framework is facilitated by the \texttt{Deblender} class, which aims to standardize the inputs and outputs of any given deblending algorithm. The input is always an instance of the \texttt{BlendBatch} class (defined in \rsec{generation}), and the outputs are structured into the \texttt{DeblendBatch} class containing the following attributes:

\begin{enumerate}
    \item \textbf{Predicted catalog}: contains detection information about each galaxy (i.e., centroid locations), as well as additional optional measurements of galaxy properties. 
    
    \item \textbf{Deblended images}: contains the reconstruction of each individually detected galaxy.

    \item \textbf{Segmentation}: contains pixel level segmentation information of the blended image, i.e., for each detected source, and for each pixel of the blended image, a value of $0$ or $1$ indicating whether the given source contributes flux to that pixel.\footnote{We opt to use this data format for image segmentation, rather than allowing are more flexible representation with values between $0$ and $1$ (perhaps corresponding to the fraction of the flux for each source), because our current deblenders and metrics would not take advantage of the additional flexibility.}
\end{enumerate}

Note that not all of these outputs are required as output of a given \texttt{Deblender} subclass. This means that our framework accommodates algorithms that only perform detection or only reconstruction, for instance.

Users are encouraged to contribute to our growing library of \btk deblenders by implementing their own algorithm as a subclass of \texttt{Deblender}. This would enable other \btk users to use their deblender and consistently compare with existing ones.
We provide several examples of ready-to-use deblenders, and provide tutorials on how this class should be structured.

Currently in \btk, the following deblenders are implemented: 
\begin{itemize}
    \item Peak-finding \citep{scikit-image2014}: Basic algorithm for finding local maxima. It takes as input a target image with one or more light sources, the minimum distance between detected peaks, the minimum intensity of peaks, and the maximum number of peaks that can be detected. It outputs a detection catalog corresponding to the detected peaks.

    \item Source Extractor \citep{bertin1996sextractor,barbary2016sep}: We use the python implementation (SEP) of this widely-used software for generating light source catalogs. In the context of \btk, the input to this function is the target image, the minimum distance between matches (in arcseconds), and the minimum pixel value for detection. This algorithm outputs a detection catalog and a segmentation, which is in turn used to provide a basic reconstruction (by extracting the pixels corresponding to the segmentation from the target image).
    
    \item \scarlet \citep{melchior2018scarlet}: Deblending algorithm using non-parametric constrained optimization, and which is integrated into the LSST software pipeline. The input includes the detections, the choice of \scarlet models be optimized for each detection, the PSF, and the noise level. The output is a deblended galaxy reconstruction for each detection. 
    
\end{itemize}
Specifically, for Source Extractor, we use the \texttt{SEP} (Source Extractor in Python) implementation \citep{barbary2016sep}, and the \texttt{peak\_local\_max} function\footnote{\url{https://scikit-image.org/docs/stable/api/skimage.feature.html\#skimage.feature.peak\_local\_max}} within \texttt{scikit-image} \citep{scikit-image2014} for the peak-finding algorithm.

\subsection{Metrics and Measurement}
\label{sec:metrics}

\btk provides a suite of metrics for evaluating deblending algorithms,
which is designed to be applied given the \texttt{BlendBatch} and \texttt{DeblendBatch} classes outputted from the previous two modules. 
However, the tutorials and documentation also demonstrate the expected input format for these metrics, so that they could also be used in isolation from the standard \btk pipeline. 
An outline of overall structure of this module is also  presented in \rfig{metrics}.

The \btk metrics are divided into three broad categories: 
\begin{itemize}
    \item \textbf{Detection}: These are metrics related to correctly counting the number of galaxies in a given image, as well as the accuracy of the predicted centroid of that galaxy. Examples of these metrics include \textit{precision}, \textit{recall}, and the \textit{efficiency matrix}.

    \item \textbf{Segmentation}: These metrics evaluate the quality of assignments of individual pixels to a corresponding identified source. An example of this metric type is \textit{intersection-over-union} (IoU).

    \item \textbf{Reconstruction}: These metrics relate to the correctness of the individual galaxy images obtained from a given galaxy blend. Examples of this metric type include the \textit{peak signal-to-noise ratio} (PSNR) \citep{ponomarenko2011psnr} and the \textit{structural similarity index} (SSIM) \citep{ssim2004}.
\end{itemize} 
We provide a full list of the currently implemented metrics in \btk in \rtab{metrics}. For each metric, we also provide the corresponding equation and a brief  description. Each of these metrics has been used previously at least once in the galaxy deblending literature.

All of the metrics implemented in \btk require information about the correspondence between detections and true sources, i.e., the \textit{matching} between the true and predicted galaxy catalog. 
In particular, the detection metrics require knowledge of which detections actually correspond to true objects, so that the number of true positives can be computed, while the other two types of metrics compare segmentations or reconstructions of galaxy images in a one-to-one basis.
Given that the matching procedure can be subject to different user choices, we provide this functionality within \btk, through the \textit{Matcher} class, whose implementation through subclasses performs this procedure. 

We currently have implemented two different algorithms to perform matching, both of which are based on the relative distances between true centroids and detections, as well as their respective counts. 
The first implemented matcher is the \texttt{PixelHungarianMatcher}, which matches galaxy centroids and detections based on their pixel value. 
This matcher is based on the Hungarian algorithm \citep{hungarian1955}, which finds an optimal weight matching in bipartite graphs. We use the implementation available as the SciPy \citep{scipy2020} function \texttt{linear\_sum\_assignment}. 
The other matcher implemented within \btk is the \texttt{SkyClosestNeighbourMatcher}, which performs matching using (ra, dec) coordinates using the \texttt{SkyCoord.match\_to\_catalog\_sky} method in Astropy \citep{astropy:2013,astropy:2018,astropy:2022}. This method internally uses KD Trees and a nearest neighbor algorithm to output matches.
Additional custom matching algorithms can be implemented in \btk creating subclasses of the \textit{Matcher} parent class.

Finally, for additional evaluation of the deblending output, we provide a set of utility functions that can compute ``measurements'' based on the simulated galaxy images generated by \btk. 
The output of these measurement functions could be used to further evaluate the reconstruction from deblenders, or to assess the challenge posed by a particular galaxy blend. 
Currently, \btk provides the following measure functions: 
\begin{itemize}
    \item \textit{Ellipticity measurement} of galaxy shapes using the KSB shear estimator implementation from the \texttt{galsim.hsm} module. 

    \item \textit{Aperture Photometry}, we provide a wrapper around the SEP aperture photometry methods to compute aperture photometry fluxes in \btk. 

    \item \textit{Blendedness} ($B$), which characterizes the degree to which the flux from a given light source overlaps with the flux of its neighbors. We use the definition from \citet{bosch2017hsc_pipeline}. 

    \item \textit{Signal-to-Noise ratio (SNR)}, which characterizes the relative brightness of the source with respect to the noise level of the image. We use the definition in \citet{arcelin2021vae}. 

\end{itemize}

\begin{table*}
\centering
\begin{tabular}{||c c c c||} 
 \hline
 \textbf{Metric} & \textbf{Type} & \textbf{Equation} & \textbf{Description} \\ [0.5ex] 
 \hline\hline
 Precision & Detection & $p = {\rm TP} / P$ & \makecell[l]{The number of detections that are matched to sources \\ (TP) divided by the total number of detections (P). \\ It represents the quality of  detections.} \\ 
 \hline
 Recall & Detection & $r = {\rm TP} / {\rm T}$ & \makecell[l]{The number of detections that are matched to sources \\ (TP) divided by the total number of  true  sources (T). \\ It represents the fraction  of sources successfully \\ detected.} \\
 \hline
 F1-score & Detection & $F_{1} = 2 \cdot (r^{-1} + p^{-1})^{-1}$ & \makecell[l]{Harmonic mean between precision $p$ and recall $r$, \\ which symmetrically represents both in one metric.} \\
 \hline
 Efficiency Matrix & Detection & $E_{ij}$ & \makecell[l]{The $i$-th, $j$-th element of this matrix is the number \\ of times that $i$ number of galaxies were detected and \\ matched in a blend containing $j$ true galaxies. } \\ 
 \hline \hline
 Intersection-over-Union & Segmentation & $\text{IoU} = \frac{S_{\text{true}} \cap S_{\text{pred}}}{S_{\text{true}} \cup S_{\text{pred}}}$ & \makecell[l]{Ratio of the intersection area between the true $S_{\rm true}$ \\ and predicted $S_{\rm pred}$ segmentation and their union area.} \\
 \hline \hline
 Mean-squared-error & Reconstruction & $\text{MSR} = \frac{1}{N} \sum_{i=0}^{N} (x^{\rm true}_i-x^{\rm pred}_i)^2$ & \makecell[l]{The mean squared residual between the true $x^{\rm true}$ and \\ the reconstructed $x^{\rm pred}$ image, where $N$ is number \\ of pixels in the images.} \\ 
 \hline
 Peak Signal-to-Noise ratio & Reconstruction & $\text{PSNR} = \log(M / \text{MSR}^2)$ & \makecell[l]{The ratio between the maximum power of the signal \\ ($M$) and the corrupting noise $(\rm MSR^{2})$.} \\ 
 \hline 
 Structural Similarity Index & Reconstruction & $\text{SSIM} = l \cdot c \cdot s$ & \makecell[l]{A more nuanced measure of similarity accounting for \\ the luminance $l$, the contrast $c$, and the structure $s$ \\ when comparing two images. (See \cite{ssim2004} \\ for details.)} \\
 \hline
\end{tabular}
\caption{\textbf{Metrics within \btk}. List of all the metrics available along with their type, equation, and a short description.}
\label{tab:metrics}
\end{table*}

\section{Tutorials and Examples}
\label{sec:tutorials}

We have included comprehensive documentation\footnote{\url{https://lsstdesc.org/BlendingToolKit/}} in \btk in order to make this package as accessible as possible.
In addition to the complete API, it contains a step-by-step user guide, instructions on how to acquire the galaxy catalogs, as well as a set of detailed tutorials with code examples. A code example demonstrating the \btk API is presented in \rapp{api}.
Tutorials are available in the \btk GitHub repository\footnote{\url{https://github.com/LSSTDESC/BlendingToolKit}} as Jupyter notebooks. The tutorials start with the basics of using \btk, and progress to advanced applications and customization of the codebase.

\subsection{List of tutorials}

In this section, we present the current list of tutorials. Note that the numbered prefix of the notebook names indicates relative complexity in increasing order.

\begin{itemize}
    \item \texttt{00-quickstart}: This tutorial is an introduction to \btk, which goes through the entire pipeline of simulating data, deblending, and evaluating results with metrics. It presents the simplest version of the \btk workflow, making it ideal for newcomers to the codebase. More details on each module and customization options are presented in tutorials that follow.
    
    \item \texttt{01-advanced-generation}: This tutorial is a deeper dive into the \texttt{generation} module (Section~\ref{sec:generation}) within \btk and showcases the different types of sampling functions, surveys, and observing condition options. 
    We also demonstrate with examples how users would write their own sampling functions, vary the input PSF, and customize the survey objects. 
    Finally, we also demonstrate how to use COSMOS Real galaxies \citep{mandelbaum2012cosmos_dataset} option within \btk to simulate galaxies with realistic morphology.
    
    \item \texttt{01-advanced-deblending}: This tutorial demonstrates usage of all the deblenders currently available within \btk (\rsec{deblend}), including how to customize their hyper-parameters, and what type of output is available from them. Finally, we also show how to run multiple deblenders on the same \btk generated data, which enables consistent comparisons between deblenders. 

    \item \texttt{01-advanced-metrics}: This tutorial shows how to use all of the metrics within the metric library of \btk (\rsec{metrics}), as well as explanations for the output, and plots on simple examples to develop intuition. 
    We also demonstrate the measure functions that are available, and how to use them on \btk data products. Finally, we explain in more detail the matching procedure in \btk and the types of matching algorithms that are available.

    \item \texttt{02-advanced-plots}: In this tutorial, we create more complex analysis plots and evaluate multiple deblenders simultaneously within the \btk framework. 
    For example, we demonstrate a comparison of two deblenders applied on the same simulated data that perform detection, and bin their precision and recall metrics by signal-to-noise (SNR). 
    We also demonstrate applying measure functions to the output of the \scarlet deblender to compute the corresponding ellipticity residuals binned by SNR.
\end{itemize}

\subsection{Example analysis plots}
\label{sec:examples}

The final tutorial \texttt{02-advanced-plots} includes a variety of useful evaluation plots involving the use of multiple deblenders and metrics. In this section, we present the plots produced in this notebook as an example of the types of analyses that \btk could facilitate. The exact definition of all the metrics applied to produce the corresponding figures in this section can be found in \rtab{metrics}.

\textbf{Dataset:} For all of the analysis plots we use the same \texttt{DrawBlendsGenerator}, although a different set of blends (random seeds) are used. 
We choose parametric bulge+disk+AGN blends from the CATSIM catalog for all examples (\rsec{generation}), where each blend can have anywhere from 0 to 10 galaxies, and the maximum distance of any source from the center of the $120 \times 120 $ pixel stamp is 15 pixels, which makes blends highly likely. 
We choose to use the LSST $r$-band survey for all experiments with corresponding parameters presented in \rapp{surveycodex-api}. Finally, we apply an AB magnitude cut of 18 on the bright end and 27 on the dim end. 

\textbf{Deblenders:} In terms of the specific settings or hyper-parameters\footnote{We emphasize that the plots presented in this sections are meant to illustrate how the codebase could be used, and not to comprehensively evaluate these deblenders. Thus, we have chosen settings for these algorithms that are reasonable for our dataset, but that might not be optimal.} of the deblenders used, these are as follows for all experiments: 
\begin{itemize}
    \item \textit{Scarlet}: We use exactly the same default settings as those in the \scarlet quick start guide\footnote{\url{https://pmelchior.github.io/scarlet/0-quickstart.html}}.

    \item \textit{SEP}: We use a \texttt{thresh} parameter of $1.5$ and \texttt{min\_area} of $5$ pixels with the \texttt{sep.extract} function\footnote{\url{https://sep.readthedocs.io/en/stable/api/sep.extract.html\#sep.extract}}, which are close to known reasonable settings for galaxy detection and photometry \citep[e.g.,][]{sanchez2021effects}. We allow SEP to measure the background of each stamp and use this for the corresponding noise variance. Otherwise, we use the default SEP parameters.

    \item \textit{Peak-finder}: We setup the \texttt{peak\_local\_max} algorithm with a \texttt{threshold} argument of $5$ sigma above the noise level, and a \texttt{min\_distance} argument equal to the PSF FWHM. These settings for the peak-finder are chosen conservatively to ensure high precision in its detections.

\end{itemize}

\begin{figure}
    \centering
    \includegraphics[width=\columnwidth]{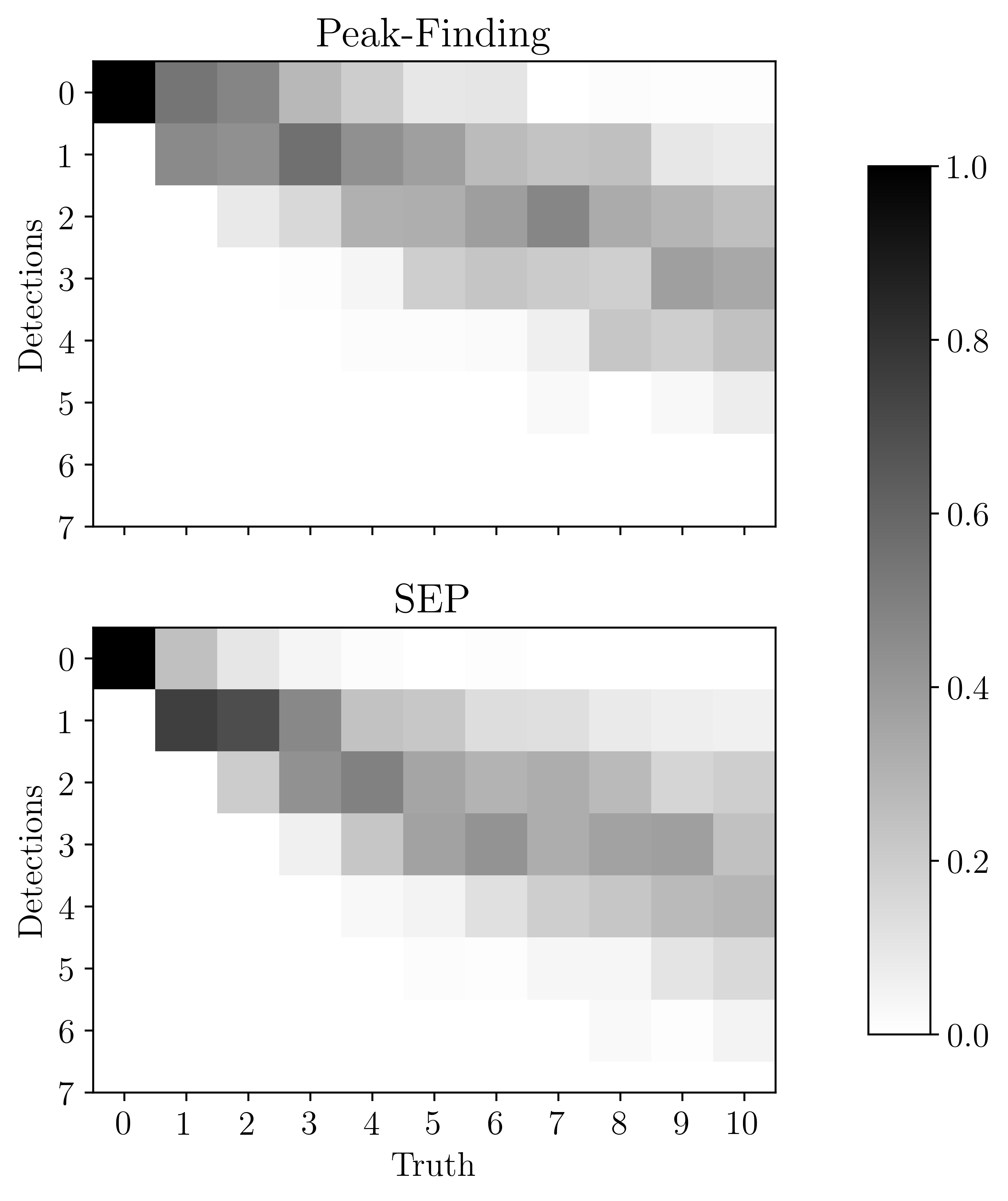}
    \caption{
        \textbf{Efficiency matrices of SEP and Peak-finding deblenders.} In this plot we show the efficiency matrices (as defined in \rtab{metrics}) for the SEP and Peak-finding deblenders implemented in \btk applied to the dataset of CATSIM blends detailed in \rsec{examples}. The rows correspond to the number of detections and the columns to the true number of sources in a blend. Each cell of each matrix represents the fraction of times that each deblender predicted the row number of detections for the column number of true sources in a blend.
        See \rsec{examples} for more details on this figure.
    }
    \label{fig:efficiency-matrix}
\end{figure}

First, we compare the detection efficiency of SourceExtractor in Python (SEP) and the peak-finding algorithm from \texttt{scikit-image}. 
Both of the efficiency matrices are displayed in \rfig{efficiency-matrix}, where the top matrix corresponds to peak-finding and the bottom to SEP. The efficiency matrices have been computed by accumulating the counts from a total of $1000$ blends, and normalizing each column by the sum of that column. 
The colorbar in the figure thus goes from $0$ (fully white) to $1$ (fully black), where $0$ for a given cell $(i,j)$ means that for blends containing $j$ galaxies, the deblender reported $i$ (matched) detections $0$ times. Similarly, a value of $1$ for a cell $(i,j)$ means that the deblender reported the correct number of (matched) detections $100\%$ of the times for blends containing $j$ galaxies. 
We can see from the efficiency matrices that SEP is overall better than the peak-finding algorithm at detecting sources in the blends we simulated, as more gray squares in the SEP efficiency matrix are closer to the diagonal. Moreover, SEP is able to detect up to $6$ sources in blends with more than $8$ galaxies whereas the peak-finding algorithm reaches a maximum of $5$ accurate detection across blends of all sizes.

\begin{figure*}
    \centering
    \includegraphics[width=0.80\textwidth]{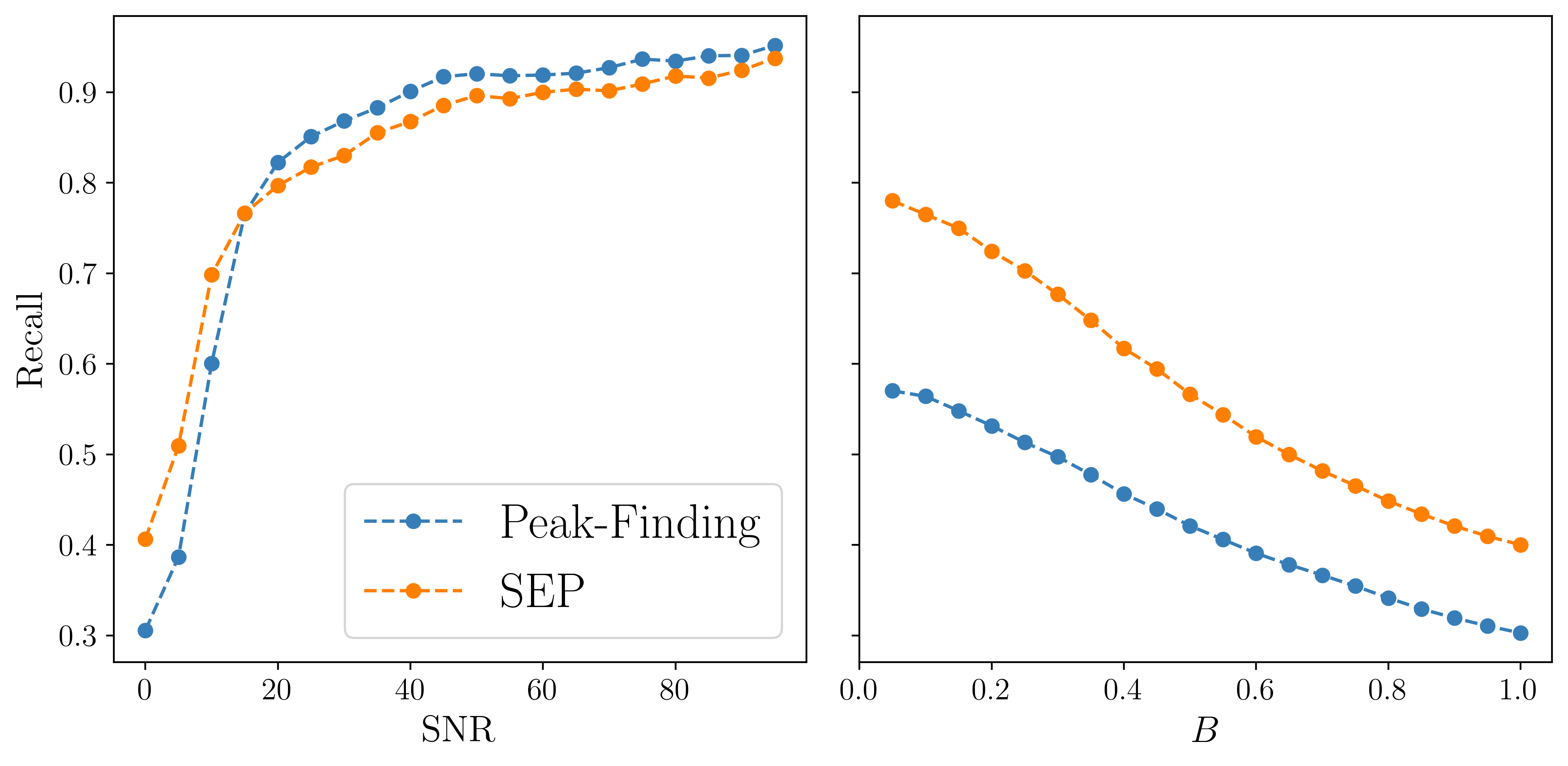}
    \caption{
    \textbf{Recall as a function of SNR and Blendedness for SEP and Peak-finding.} 
    In this figure, we compare the achieved recall (as defined in \rtab{metrics}) by the peak-finding (blue) and SEP (orange) deblenders as a function of the minimum SNR and maximum blendedness ($B$) of true galaxies applied to the same CATSIM blend dataset as Figure~\ref{fig:efficiency-matrix}.
    See \rsec{examples} for more details on this figure.
    }
    \label{fig:recall}
\end{figure*}

Next, in \rfig{recall} we show the recall of these two detection algorithms as a function of the true minimum SNR and maximum blendedness $B$ of the galaxies in a given blend. The recall is computed by taking the number of detections that are matched with true galaxies whose SNR or blendedness satisfies the threshold, and dividing by the number of true galaxies satisfying the threshold.
In the left plot, we see that as the minimum SNR threshold is increased, the recall increases for both deblenders. This is expected as we are excluding fainter objects as the threshold increases. 
However, we do not expect to achieve perfect recall even at very high SNRs due to blending. This is indeed what we see for both algorithms, as they both reach only approximately $0.9$ recall at the brightest threshold.
In terms of the relative performance between SEP and the peak-finder, SEP achieves significantly better recall in the low-SNR bins and the peak-finder achieves slightly better recall in the high-SNR bins. This is reasonable, as Source Extractor's thresholding approach is more suitable for detecting faint sources, where as peak-finding is more appropriate for bright, non-diffused light sources.
In the right plot, we see that both methods achieve their highest recall when restricted to galaxies with low blendedness, and their recall smoothly decreases as we include more blended objects. We see that SEP performs significantly better than the peak-finder with our settings for all blendedness thresholds. 
Note that we do not include the corresponding precision plots as both algorithms achieve an overall precision close to $99\%$ for our dataset.

\begin{figure*}
    \centering
    \includegraphics[width=0.95\textwidth]{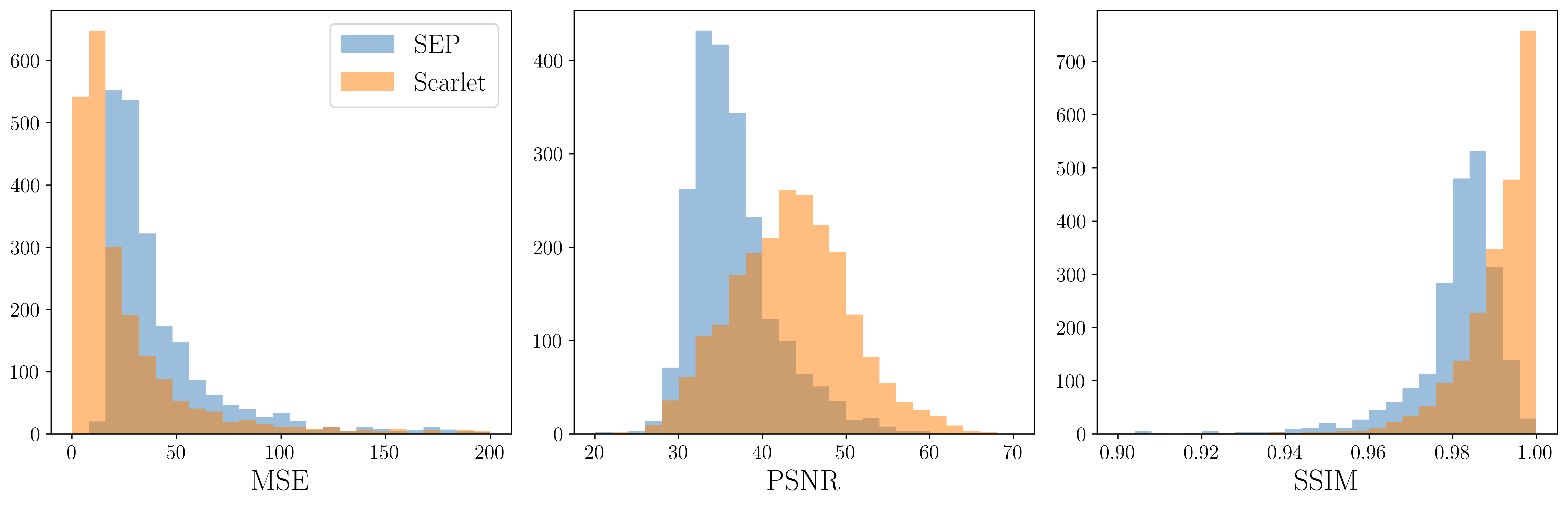}
    \caption{
        \textbf{Reconstruction metrics of SEP and Scarlet.} In this figure we show three histogram plots for each of the reconstruction metrics available in \btk: mean-squared error (MSE), peak SNR (PSNR), and structural similarity index (SSIM), as defined in \rtab{metrics} for the dataset of CATSIM blends detailed in \rsec{examples}.
        The light blue histogram in each plot corresponds to the metrics evaluated on reconstructions of galaxies from the SEP deblender, whereas the orange histogram corresponds to the \scarlet deblender. Each count of each histogram corresponds to the given metric computed from the reconstruction of either SEP or \scarlet of a given detection that was matched to a true galaxy centroid. See \rsec{examples} for more details on this figure.
    }
    \label{fig:reconstruction-histograms}
\end{figure*}

Third, in \rfig{reconstruction-histograms} we show three histograms corresponding to the reconstruction metrics available within \btk (see \rtab{metrics}), comparing SEP and the \scarlet deblender.
Each of the histograms is composed of the metric applied to the reconstructions of true galaxies that were detected and matched by the SEP detection algorithm (since \scarlet does not perform detection). The SEP's reconstruction of a given galaxy is obtained by multiplying the pixel-wise segmentation outputted by SEP with the galaxy blend image. 
Overall, we see from the three histograms that \scarlet's reconstructions are higher quality than SEP's reconstructions. For instance, the Mean-Squared Error (MSE) between \scarlet's reconstructions and the true noiseless individual galaxy images are closer to zero on average than SEP's reconstructions are.\footnote{An important caveat is that the MSE metric improves as the size of the stamp grows bigger regardless of the quality of the reconstruction.} Moreover, the structure similarity index is closer to 1 on average for \scarlet, indicating higher visual similarity than SEP's reconstructions. Both of these results are expected as \scarlet uses a more sophisticated constrained matrix factorization procedure to deblend light sources than SEP.

\begin{figure*}
    \centering
    \includegraphics[width=0.99\textwidth]{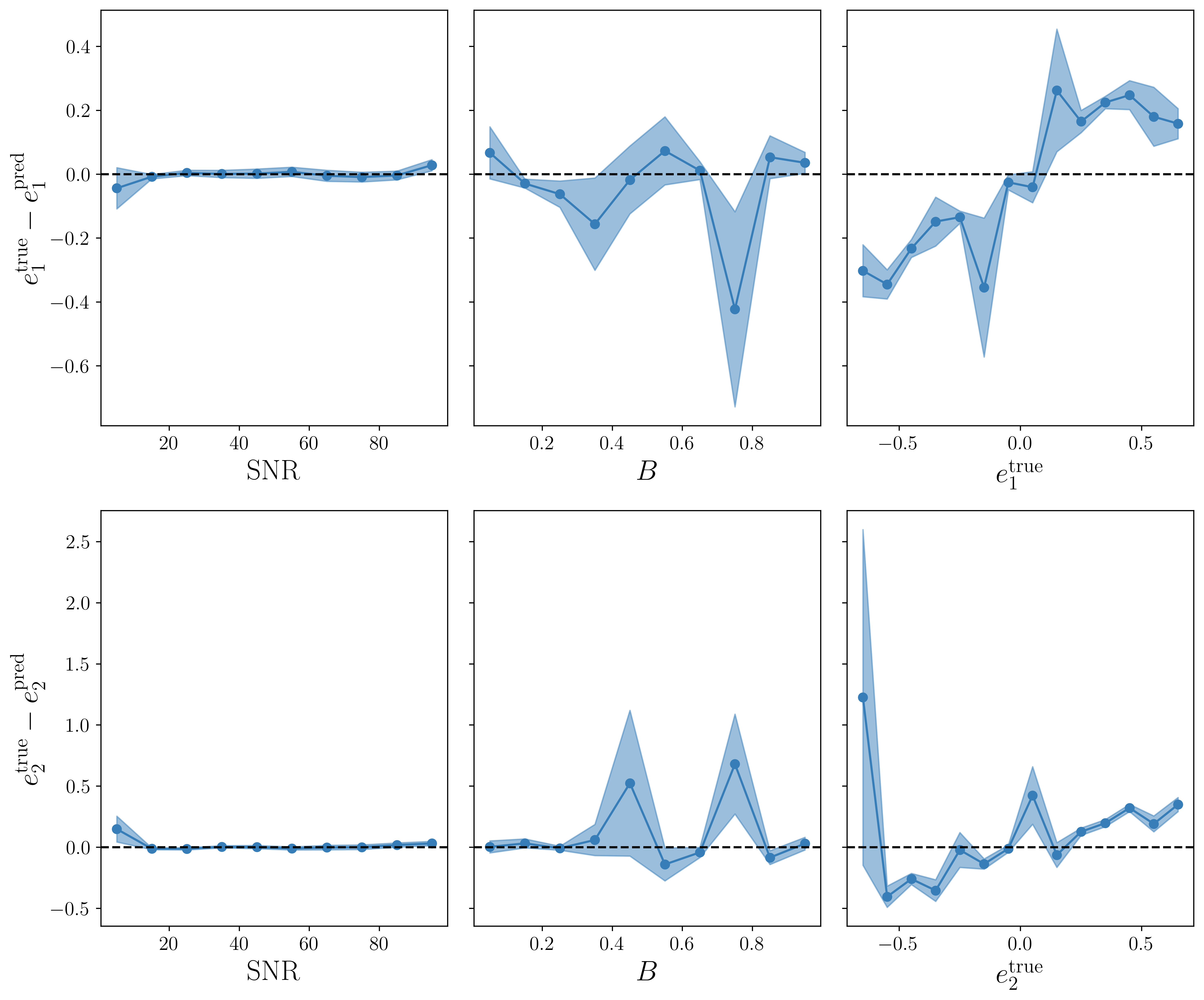}
    \caption{
        \textbf{Ellipticity residuals of Scarlet deblender.} In this figure we show the HSM ellipticity residuals between the reconstructed individual galaxies from \scarlet and the true isolated galaxy images binned by signal-to-noise ratio (SNR), blendedness ($B$), and true measured ellipticity ($e_{1}^{\rm true}$). The curve refers to the mean value for that bin and shaded region corresponds to the error on the mean.
        We emphasize that the dataset of galaxy blends used for this example is strongly blended, so that these plots might not be representative of the performance of \scarlet on future data. 
        See \rsec{examples} for more details on this figure.
    }
    \label{fig:ellipticity-residuals}
\end{figure*}

In \rfig{ellipticity-residuals} we show the residuals between HSM ellipticity\footnote{Specifically, we are using the `KSB' method within the HSM module in \galsim \citep{hsm2003}.} measured on reconstructions of individual galaxies from \scarlet and on the true individual galaxy images binned by SNR, blendedness ($B$), and HSM ellipticity  measured on the true galaxy images ($e_{1,2}^{\rm true}$). 
For this figure, we condition \scarlet on the true centroids of every galaxy in the blend, thus ignoring the effect of unrecognized blends.
In each figure, the curve and points correspond to the mean value of each bin, and the shaded bands correspond to the error on the mean for that bin ($\sigma / \sqrt{N_{\rm bin}}$, where $\sigma$ is the residual scatter in the bin and $N_{\rm bin}$ is the number of galaxies in that bin). 
For the ellipticity residual as a function of SNR, we see essentially unbiased reconstructed ellipticity with scatter increasing for decreasing SNR. For blendedness, we see large scatters in the mean residual for some of the bins, which suggests that a couple of outliers are driving the large deviations from mean. Finally, for the functions binned by ellipticity, we see a significant dependence on the mean residual as a function of true ellipticity. 

We note that the galaxy blend dataset used in \rfig{ellipticity-residuals} is very strongly blended, and, as stated above, we run \scarlet using the default settings from its basic tutorial. Therefore, this figure might not necessarily be representative of the performance of \scarlet on future survey data. Additionally, the mean of HSM shape measurements is designed to be unbiased (to some degree) only in the case of a small shear applied to a galaxy population with an isotropic distribution of ellipticities. Thus, we do not expect this residual mean to be unbiased as a function of ellipticity even in the case of no blending. 

\begin{figure*}
    \centering
    \includegraphics[width=0.90\textwidth]{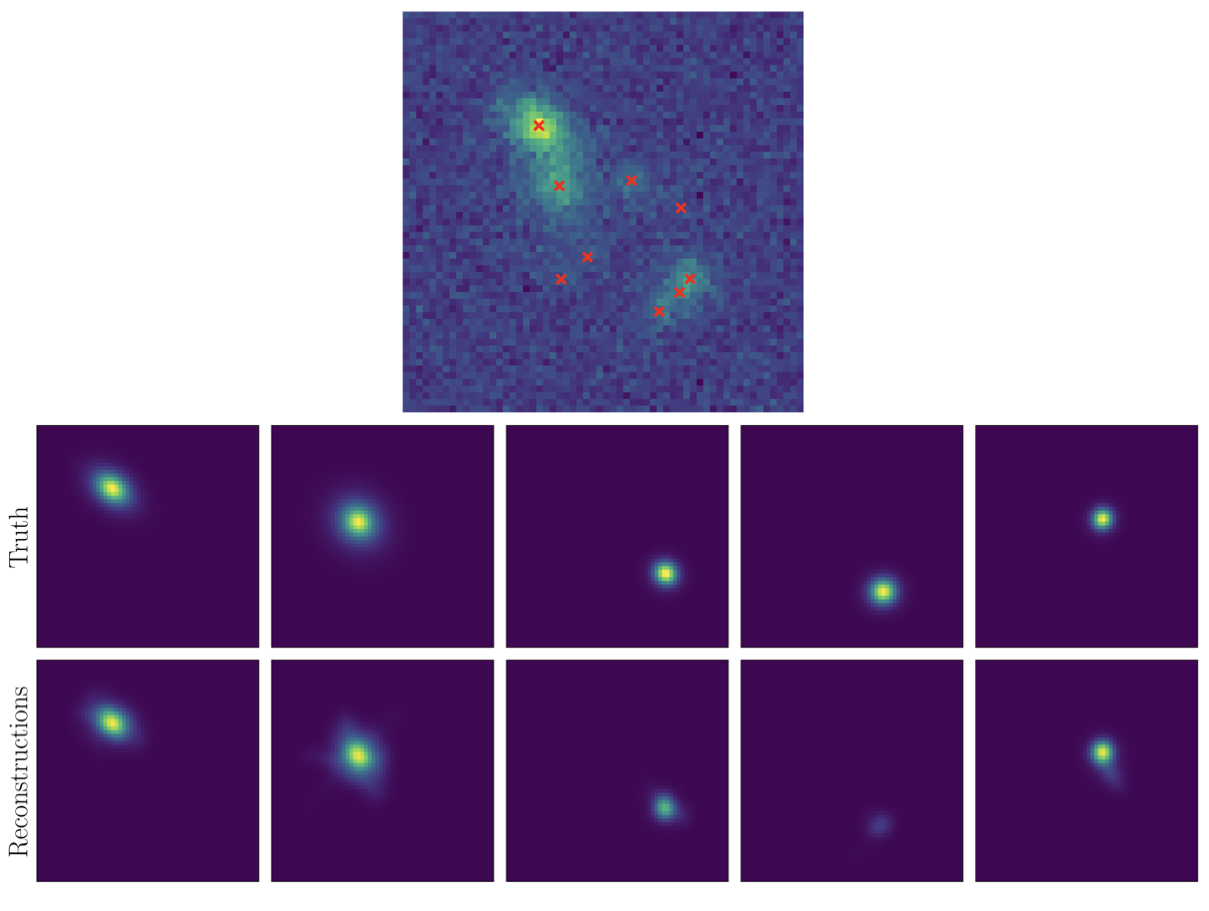}
    \caption{
        \textbf{Example of Scarlet Reconstructions.} In this figure we showcase an example of a galaxy blend image (top, middle) with the centroid of each galaxy marked in red, and the true (noiseless) image of each individual galaxy (top row) along with the respective \scarlet reconstruction (bottom row). Each of the galaxies are ordered by total flux from left to right, and their positions are the same as in the original blend.
    }
    \label{fig:reconstruction-example}
\end{figure*}

Lastly, in \rfig{reconstruction-example} we show an example of a galaxy blend, the true image of each member galaxy, and the respective reconstructions from \scarlet (using the true galaxy centroid as input). We see how the reconstruction of a given individual galaxy can suffer from significant artifacts when it is highly blended or when its SNR is low. 

\section{Summary}
\label{sec:summary}

In this paper, we have introduced \btk, an open source package that provides customizable galaxy blends, a standardized framework for deblenders, and a library of deblending metrics that enables self-consistent comparisons.
We developed \btk in the context of the Blending Working Group in LSST DESC, in order to further enable the development of deblenders within the collaboration. 
However, we anticipate that \btk will be useful to anyone interested in analyzing the systematic errors that blending induces in measurements of astronomical sources. 

\btk also provides many useful subroutines for understanding galaxy blends and enables an end-to-end analysis. For instance, we provide a \texttt{match} module that provides various (customizable) ways of assigning true objects to detections for evaluating the \textit{accuracy} of the reconstruction. This tends to be a subtle step when evaluating deblenders that is performed differently by different studies. Standardizing this step with its own module within \btk allows for more principled comparisons between deblenders.
Additionally, \btk provides several utility functions for extracting measured quantities of individual galaxies or blends that are of interest for blending analyses. These quantities include aperture photometry flux and blendedness. Standardized functions that extract these quantities given \btk output facilitates downstream analyses. 

As described in more detail in \rsec{development}, \btk is an open source package hosted on GitHub, and we strive to maintain a culture of open collaboration and development. 
Anyone interested in galaxy blending is invited to contribute to this package. 
We have comprehensively tested and documented the software in preparation of the first formal release of the package, v1.0, which accompanies this publication. Finally, we hope that this package and its tutorial suite serve as a pedagogical resource for those interested in learning the basics of blending analysis and deblenders.

The simulations generated by \btk are actively used by the LSST DESC community to test various deblenders in development \citep{merz2023deepdisc,biswas2024madness}, and efforts are underway to integrate some of these deblenders into the \texttt{deblender} module within \btk.
The development of \btk continues after this first release of the software and some future directions include: improving the overall realism of galaxy simulations, expanding the metrics library to allow more types of science analysis beyond those involving galaxy fluxes and shapes, and incorporating more modern galaxy catalogs such as those from the DC2 simulation \citep{lsst2021dc2}.

\section{Author Contributions} \label{sec:contributions}

IM was the lead developer and maintainer of the codebase. 
IM, AT, and TS wrote the paper, created and polished figures, and developed tutorials. 
IM and TS documented the codebase, developed unit tests, and were co-leads of the \btk team within LSST-DESC.
AG gave extensive and detailed guidance on codebase development and structure.
AG, CA, EA, and CR were mentors for the primary codebase contributors. 
AB was the lead developer of \surveycodex, and MP and HM contributed to this codebase.
IM, AT, TS, AG, MP, CA, BB, CD, RJ, AIM, HM, and TZ brainstormed and planned \btk codebase development. 
IM, AT, TS, AB, MP, PA, BB, CD, RJ, SK, and TZ contributed to the codebase. 
IM, AT, TS, AG, MP, PA, BB, GM, and TZ identified and fixed bugs within the codebase. 
IM, AT, TS, and AB performed code reviews for the codebase. 
MP incorporated ``Real'' Galsim galaxies into \btk.
AG, AB, CA, JB, PB, RJ, and GM gave paper suggestions, edits, and helped improved paper structure.
EA, PB, CD, SK, and CR were involved in the initial planning and conception of \btk. 
JB and TZ managed the working-group in DESC that the BTK team was part of.
SK developed the first draft of the codebase. 

\section{Acknowledgements} \label{sec:acknowledgements}

This paper has undergone internal review in the LSST Dark Energy Science Collaboration. We are grateful to our internal reviewers, Mike Jarvis and Shuang Liang, for their thoughtful feedback.

IM and CA acknowledge support from DOE grant DE-SC009193. IM acknowledges the support of the Special Interest Group on High Performance Computing (SIGHPC) Computational and Data Science Fellowship. IM acknowledges support from the Michigan Institute for Computational Discovery and Engineering (MICDE) Graduate Fellowship. IM acknowledges support from the LSSTC Enabling Science Award. IM acknowledges support from the Leinweber Center for Theoretical Physics Summer Fellowship.
AIM acknowledges the support of Schmidt Sciences.
MP and HM acknowledge supports from Japan Society for the Promotion of Science (JSPS) KAKENHI Grant Numbers JP20H01932 and JP23H00108, and Japan Science and Technology Agency (JST) CREST Grant Number JPMHCR1414 and FOREST Program Grant Number JPMJFR2137. GM acknowledges support from NSF grant AST-2308174 and LSST-DA through grant 2023-SFF-LFI-03-Liu. 
BB acknowledges support from Agence Nationale de la Recherche grant ANR-19-CE23-0024 - AstroDeep, and the European Union's Horizon 2020 research and innovation program under the Marie Sk\l{}odowska-Curie grant agreement No 945304 - COFUND AI4theSciences hosted by PSL University.

The DESC acknowledges ongoing support from the Institut National de 
Physique Nucl\'eaire et de Physique des Particules in France; the 
Science \& Technology Facilities Council in the United Kingdom; and the
Department of Energy, the National Science Foundation, and the LSST 
Corporation in the United States.  DESC uses resources of the IN2P3 
Computing Center (CC-IN2P3--Lyon/Villeurbanne - France) funded by the 
Centre National de la Recherche Scientifique; the National Energy 
Research Scientific Computing Center, a DOE Office of Science User 
Facility supported by the Office of Science of the U.S.\ Department of
Energy under Contract No.\ DE-AC02-05CH11231; STFC DiRAC HPC Facilities, 
funded by UK BEIS National E-infrastructure capital grants; and the UK 
particle physics grid, supported by the GridPP Collaboration.  This 
work was performed in part under DOE Contract DE-AC02-76SF00515.

We also acknowledge the use of \texttt{numpy} \citep{numpy2020}, \texttt{scipy} \citep{scipy2020}, \texttt{scikit-learn} \citep{scikit-learn}, \texttt{scikit-image} \citep{scikit-image2014}, \astropy \citep{astropy:2013,astropy:2018,astropy:2022}, \texttt{matplotlib} \citep{matplotlib2007}, \texttt{fast3tree}\footnote{\url{https://github.com/yymao/fast3tree}}, \texttt{tqdm}\footnote{\url{https://github.com/tqdm/tqdm}}, \texttt{h5py}\footnote{\url{https://github.com/h5py/h5py}}, and \galsim \citep{galsim2015}.

\section*{Data Availability}
 
The BlendingToolKit (\btk) is a publicly available package on GitHub at the following link: \url{https://github.com/LSSTDESC/BlendingToolKit}. Our software is also publicly available in Zenodo at \url{https://zenodo.org/records/13351808}. To install the software, start with the documentation at
 \url{https://lsstdesc.org/BlendingToolKit}.

\bibliographystyle{aasjournal}
\bibliography{main}

\begin{appendix}
\section{API Demonstration}
\label{app:api}

In what follows we illustrate the \btk API to generate blended images, run a deblender on these, and evaluate the performance of the deblender using detection metrics within \btk. This code example is a condensed version of our ``quickstart'' tutorial (see \rsec{tutorials}).
\begin{python}
import btk

# setup CATSIM catalog
catalog_name = "../data/input_catalog.fits"
catalog = btk.catalog.CatsimCatalog.from_file(catalog_name)

# setup survey parameters
survey = btk.survey.get_surveys("LSST")

# setup sampling function
# this function determines how to organize galaxies in catalog into blends
stamp_size = 24.0
sampling_function = btk.sampling_functions.DefaultSampling(
    catalog=catalog, max_number=5, max_mag=25.3, stamp_size=stamp_size
)

# setup generator to create batches of blends
batch_size = 100
draw_generator = btk.draw_blends.CatsimGenerator(
    catalog, sampling_function, survey, batch_size, stamp_size
)

# get batch of blends
blend_batch = next(draw_generator)

# setup deblender (we use SEP in this case)
deblender = btk.deblend.SepSingleBand(
    max_n_sources=5,
    use_band=2 # measure on 'r' band
)

# run deblender on generated blend images (all batches)
deblend_batch = deblender(blend_batch)

# setup matcher
matcher = btk.match.PixelHungarianMatcher(pixel_max_sep=5.0) # maximum separation in pixels for matching

# match true and predicted catalogs
truth_catalogs = blend_batch.catalog_list
pred_catalogs = deblend_batch.catalog_list
matching = matcher(truth_catalogs, pred_catalogs) # object with matching information

# compute detection performance on this batch
recall = btk.metrics.detection.Recall(batch_size)
precision = btk.metrics.detection.Precision(batch_size)
print("Recall: ", recall(matching.tp, matching.t, matching.p))
print("Precision: ", precision(matching.tp, matching.t, matching.p))
\end{python}

\section{\surveycodex}
\label{app:surveycodex}

\btk is made to simulate blending scenes for a variety of galaxy surveys. As such, it relies on a careful bookkeeping of the survey parameters, which are usually found scattered among various papers, websites, or code repositories, sometimes without the unit or a proper reference.

In order to improve the consistency of the simulator and prevent the occurrence of bugs in the blending scenes produced for various surveys, a subgroup of \btk developers decided to develop \surveycodex\footnote{\url{https://github.com/LSSTDESC/surveycodex}}, a tiny Python library with referenced survey parameters from the current main galaxy surveys. Consistency was achieved by associating a unit to each of these parameters through the use of \texttt{astropy.units} \citep{astropy:2013,astropy:2018,astropy:2022}. 

The available surveys in \surveycodex for the 1.0 \btk release are given in \rtab{surveycodex_surveys}. An example of command-line output from \surveycodex for the LSST survey is given in Appendix~\ref{app:surveycodex-api}. 

\subsection{Description}

The library contains parameters describing the telescope (e.g., \texttt{mirror\_diameter}), the camera (e.g., \texttt{pixel\_scale} or \texttt{gain}), and the characteristics of each of the filters (e.g., \texttt{psf\_fwhm}, \texttt{effective\_wavelength}, etc.). 
Most values have been sourced from official documents or websites, and some like the filters \texttt{effective\_wavelength} or the \texttt{zeropoint} have been computed through the \texttt{speclite}\footnote{\url{https://github.com/desihub/speclite}} package that references the filter responses of most of these surveys. A description of each parameter it holds, carefully sourced or checked with specialists from each survey, can be found in the online documentation\footnote{\url{https://lsstdesc.org/surveycodex/}}.

\surveycodex also incorporates basic utility functions, such as the computation of fluxes in electron counts from a given AB magnitude, or the mean sky level.
The former is used within \btk to compute galaxy fluxes from magnitudes in the galaxy catalog and follows the usual convention
\begin{equation}
    F = T \cdot 10^{-0.4(m-Z)},
\end{equation}
where $F$ is the flux, $T$ the exposure time, $Z$ the zeropoint, and $m$ the corresponding magnitude.

\btk includes its own \texttt{Survey} class (\rsec{generation}) that directly inherits and contains the information provided by the \surveycodex library. 

\begingroup

\setlength{\tabcolsep}{10pt} 
\renewcommand{\arraystretch}{1.4} 
\begin{table}[h]
    \centering
    \begin{tabular}{c|c|c|c}
        \multicolumn{4}{c}{\textbf{\surveycodex v1.1 available entries}} \\
        \hline 
        \textbf{Survey name} & \textbf{Telescope} & \textbf{Instrument} & \textbf{Reference} \\
        \hline
        CFHTLS & CFHT & Megacam & \citet{cuillandre2006cfht} \\
        COSMOS & HST & ACS & \citet{koekemoer2007cosmos} \\
        DES & Victor M. Blanco & DECam & \citet{des2016des}  \\
        Euclid\_VIS & Euclid & VIS & \citet{clampin2012euclid} \\
        HSC & Subaru & HSC & \citet{hiroaki2018hsc} \\
        LSST & Simonyi & LSST camera & \citet{ivezic2019lsst} \\
    \end{tabular}
    \caption{
        Supported surveys in \surveycodex and \btk for the Version 1.0 \btk release.
        }
    \label{tab:surveycodex_surveys}
\end{table}

\endgroup

\subsection{\surveycodex command-line example} \label{app:surveycodex-api}

The following is a reduced example output from the command line utility of the \surveycodex package. This output demonstrates the information accessible within \surveycodex about the LSST survey. After the parameter values, a table is shown with references for each parameter. The number of filters and the reference table at the bottom have been shortened for display purposes in this paper.

\footnotesize{
\begin{verbatim}
$ surveycodex -s LSST --ref --rich

----------------- LSST -------------------
Legacy Survey of Space and Time (LSST)
done with the Simonyi survey telescope
and the LSST camera

Pixel scale:       0.200 arcsec
Mirror diameter:   8.36 m
Effective area:    33.34 m2
Obscuration:       0.39
Gain:              1.00 electron / adu
Zeropoint airmass: 1.2
Available filters: u, g, r, i, z, y

Filter info:                                
-- u ----------------------------------------
    PSF FWHM:             0.90 arcsec
    Zeropoint:            26.40 mag
    Sky brightness:       22.99 mag / arcsec2
    Full exposure time:   1680 s
    Effective wavelength: 368.1 nm
-----------------------------------------------
-- g ----------------------------------------
    PSF FWHM:             0.86 arcsec        
    Zeropoint:            28.26 mag          
    Sky brightness:       22.26 mag / arcsec2
    Full exposure time:   2400 s             
    Effective wavelength: 486.4 nm           
-----------------------------------------------
                        ...

                                                            LSST references
| Parameter            | Source                                                              | Comments                    |
| -------------------- | ------------------------------------------------------------------- | --------------------------- |
| pixel_scale          | https://www.lsst.org/about/camera/features                          | See Technical details.      |
| gain                 | https://github.com/LSSTDESC/imSim/blob/main/imsim/stamp.py#L416     |                             |
| mirror_diameter      | https://www.lsst.org/about/tel-site/optical_design                  |                             |
| zeropoint_airmass    | https://speclite.readthedocs.io/en/latest/filters.html#lsst-filters |                             |
| sky_brightness       | https://arxiv.org/abs/0805.2366                                     | (v5) See Table 2, first row |
|         ...          |                            ...                                      |            ...              |

\end{verbatim}
}

\end{appendix}


\end{document}